\documentclass[lettersize,journal]{IEEEtran}

\usepackage{setspace}
\usepackage{amsmath,amsfonts}
\usepackage{algorithm}
\usepackage{array}
\usepackage{textcomp}
\usepackage{stfloats}
\usepackage[hyphens]{url}
\usepackage{verbatim}
\usepackage{graphicx}
\usepackage{bm}
\usepackage{pgfplots}
\usepackage{algorithm}
\usepackage{algpseudocode}
\usepackage{listings}
\usepackage{xcolor} %Optional: for customizing colors
\definecolor{mygreen}{rgb}{0,0.6,0}
\usepackage{array}  % for better column definitions
\usepackage{paralist}
\usepackage{colortbl}
\usepackage{amssymb}
\usepackage{pifont}
\usepackage{amsmath} 
\usepackage{graphicx}
\usepackage{booktabs}
\usepackage{hyperref}
\usepackage{makecell}
\usepackage{breakurl}
\usepgfplotslibrary{groupplots}
\usepackage{orcidlink}

\usepackage{colortbl}
\usepackage{xcolor}
\definecolor{graynum}{gray}{0.5}

\usepackage{amssymb}  % For \checkmark
\usepackage{tikz}
\usepackage{diagbox}  % For diagonal lines in table cells
\usepackage{adjustbox} % For adjusting position of the table
\usepackage{verbatim}
\usepackage{caption}
\usepackage{rotating}  % For rotating text
\usepackage{multirow}
\usepackage{makecell}  % For better cell formatting
\usepackage{arydshln}

\usepackage{pgffor}
\usepackage{pgfplots}
\usepackage{caption}
\usepackage{subcaption} 
\usepackage{graphicx}

\newcolumntype{C}[1]{>{\centering\arraybackslash}m{#1}}

\usepackage{graphicx} % Include any packages you might need
\usepackage{caption}
\usepackage{rotating}  % For rotating text
\usepackage{multirow}
\usepackage{makecell}  % For better cell formatting
\usepackage{amssymb}  % For \checkmark
\usepackage{array}
\usepackage{wrapfig}
\newcolumntype{P}{>{\hspace{0.3cm}}p{0.9cm}}
\newcolumntype{C}{>{\centering\arraybackslash}m{5cm}}  % Category column
\newcolumntype{X}{>{\centering\arraybackslash}m{4cm}}  % Attack column
\newcolumntype{B}{>{\centering\arraybackslash}p{0.9cm}}  % Attack column
\usepackage{pgfplots}
\usepackage[dvipsnames]{xcolor}

\newcommand{\scheme}{\textsc{FeatureBleed}}
\lstdefinestyle{boxed}{
    frame=single, % Adds a frame around the code
    xleftmargin=3pt,
    xrightmargin=3pt,
    aboveskip=4pt,
    belowskip=0pt,
}

\lstdefinestyle{nonboxed}{
    frame=none, % No frame around the code
    xleftmargin=0pt, % Adjust margins
    xrightmargin=0pt,
    aboveskip=4pt,
    belowskip=0pt,
}

\usepackage{graphicx}

\begin{document}

%\title{A Novel Side-Channel Vulnerability in Intel AMX}

% \IEEEpubid{0000-0000~\copyright~2024 IEEE}
\title{\scheme{}: Inferring Private %Retrieval 
Enriched Attributes From Sparsity-Optimized AI Accelerators}

 % %\input{Shorts/Intro-short}
 % \author{Darsh Asher, Farshad Dizani, Joshua Kalyanapu, Rosario Cammarota, Aydin Aysu, Samira Mirbagher Ajorpaz
 % }

 \author{
Darsh Asher\,\orcidlink{0009-0007-3325-0224},
Farshad Dizani\,\orcidlink{0009-0005-8136-5320},
Joshua Kalyanapu\,\orcidlink{0009-0002-3186-5069},
Rosario Cammarota\,\orcidlink{0000-0002-2965-8987},
Aydin Aysu\,\orcidlink{0000-0002-5530-8710},
Samira Mirbagher Ajorpaz\,\orcidlink{0009-0008-4997-5980}
}

%  \thanks{Manuscript received November 4, 2023; revised February 28, 2024.}
% \thanks{The next few paragraphs should contain 
% the authors' current affiliations, including current address and e-mail. For 
% example, First A. Author is with the National Institute of Standards and 
% Technology, Boulder, CO 80305 USA (e-mail: author@boulder.nist.gov). }
% \thanks{Second B. Author and Third C. Author are with Rice University, Houston, TX 77005 USA. (e-mail: authorB@lamar.colostate.edu, authorC@abc.com).}
% \markboth{IEEE Journal of \LaTeX\ Class,~Vol.~12, No.~6, February~2024}%

 \maketitle

\begin{spacing} {0.962} %{0.98}

\begin{abstract}
Backend enrichment is now widely deployed in sensitive domains such as product recommendation pipelines, healthcare, and finance, where models are trained on confidential data and retrieve private features whose values influence inference behavior while remaining hidden from the API caller. This paper presents the first \emph{hardware-level backend retrieval data-stealing attack}, showing that accelerator optimizations designed for performance can directly undermine data confidentiality and bypass state-of-the-art privacy defenses.

Our attack, \scheme{}, exploits {\emph{zero-skipping}} in AI accelerators to infer private {backend-retrieved features} solely through end-to-end timing, without relying on power analysis, DVFS manipulation, or shared-cache side channels. We evaluate \scheme{} on three datasets spanning medical and non-medical domains—Texas-100X (clinical records), OrganAMNIST (medical imaging), and Census-19 (socioeconomic data). We further evaluate \scheme{} across three hardware backends (Intel AVX, Intel AMX, and NVIDIA A100) and three model architectures (DNNs, CNNs, and hybrid CNN--MLP pipelines), demonstrating that the leakage generalizes across CPU and GPU accelerators, data modalities, and application domains, with an adversarial advantage of up to \textbf{98.87} pp.

Finally, we identify the root cause of the leakage as sparsity-driven zero-skipping in modern hardware. We quantify the privacy--performance--power trade-off: disabling zero-skipping increases Intel AMX’s per-operation energy by up to \textbf{25\%} and incurs \textbf{100\%} performance overhead. %\textcolor{blue}
{We propose a padding-based defense that masks timing leakage by equalizing responses to the worst-case execution time, achieving protection with only \textbf{7.24\%} average performance overhead and no additional power cost.}
\end{abstract}

\begin{IEEEkeywords}
Backend enrichment, Feature Store ,Activation sparsity, Zero-Skipping,
AI accelerators, Timing leakage, Machine Learning Privacy
\end{IEEEkeywords}

\vspace{-1em}
\section{Introduction}
Modern machine learning (ML) services increasingly operate in settings where the system must make real-time decisions among a large number of possible outcomes, such as selecting content, assessing risk, or generating clinical recommendations. In these settings, the input to inference is rarely limited to what the user explicitly provides. Instead, accurate decisions require \textit{additional contextual information} such as historical behavior, recent activity, or authoritative records that already exist within the service, are private, and must be \textit{retrieved} and \textit{enriched} for the ML network at request time; %this 
a widely used example of such systems is 
%called a 
Feature Store~\cite{ML_feature_stores, feast_feature_store_2025, de2024hopsworks}. %. Examples of feature stores include Feast~\cite{feast_feature_store_2025} and Hopsworks~\cite{de2024hopsworks}.

%In large-scale online inference, predictions therefore cannot be computed from the request alone and must be \textit{enriched} with backend data in real time. 

%Feature stores~\cite{uber2024predictive_to_generative}
%\textcolor{red}{cite the VLDB Managing ML pipelines paper here}
%—such as 
Uber Palette~\cite{uber2024predictive_to_generative} within the Michelangelo platform addresses this by pre-computing features offline, %maintaining their latest values online, 
and retrieving them during inference. 
%with millisecond latency, even at massive scale. 
For example, when a rider requests a trip or food delivery, the request contains only basic context (e.g., location, destination, and an entity identifier), while the ML prediction depends on \textit{private}, continuously updated features such as driver acceptance rates, past trip durations, common routes, or restaurant preparation-time statistics. These features are shared across many models, evolve over time, and cannot be re-computed or supplied by the client at inference time. This backend enrichment step is essential for real-time ML systems; without it, predictions would be inaccurate, incur unacceptable latency, or require exposing sensitive internal data to callers.

%From a threat-model perspective, the caller is a legitimate API user who can issue standard requests and observe end-to-end latency, but cannot see or modify the backend features themselves. Although these retrieved attributes remain private and are never exposed through the API, they directly influence the model’s internal computation—specifically activation patterns and sparsity—which in turn affects execution time on modern accelerators. This creates a realistic setting in which sensitive backend attributes, introduced only during inference via feature retrieval, can impact observable timing behavior and become susceptible to leakage without requiring privileged access, co-location, or control over the features.

Recommendation and personalization systems such as NVIDIA’s latest real-time recommendation pipelines
~\cite{nvidia2023featurestore}
or Microsoft Copilot %~\cite{stratton2024introduction} 
also exemplify variants of this design. Large platforms must choose a small set of relevant items from millions of candidates under strict latency constraints. %Random or uniform ordering is ineffective, and users cannot reasonably supply the rich and rapidly changing signals required for personalization. As a result, m
Modern recommender systems retrieve user- and item-specific features from backend storage during inference and use them internally to rank content. This server-side enrichment is essential for \textit{relevance}, \textit{freshness}, and \textit{robustness}, and is a core component of production feeds, ads, and search ranking systems.

A similar architecture underlies credit scoring, underwriting, and fraud detection systems operated by Visa, Mastercard, and FICO, as well as large-scale product recommendation and personalization pipelines deployed by Alibaba, Amazon, and Shopify.
% A similar architecture underlies credit \textcolor{red}{name products for each example paragraphs in the intro without citing since we dont' have space,} underwriting and fraud detection systems. 
These systems must make fast, high-stakes decisions based on sensitive and authoritative information such as transaction history, repayment behavior, device signals, or recent anomalies. Such information cannot be safely or reliably provided by users, as it is subject to manipulation, omission, or staleness. Consequently, risk and underwriting models retrieve features from \textit{trusted backend sources} at decision time and apply \textit{predictive ML models} to estimate default risk or fraud likelihood. Without backend retrieval, these systems would be inaccurate and insecure.
%, and incompatible with regulatory requirements.
A clinical decision support systems %assist %providers by incorporating up-to-date patient data%—such as laboratory results, vital signs, medications, and prior diagnoses—that are stored in electronic health records and cannot be reliably supplied by patients or client applications. 
and insurance decision systems similarly interact with controlled decision-support services that retrieve authorized summaries and coded records %(e.g., procedures, utilization history, or risk indicators) 
to assess eligibility or coverage, without exposing full clinical records to the insurer. In both settings, the entity issuing the query has legitimate access to the decision-support interface but does not observe the sensitive attributes used internally by the model. 
Retrieving these attributes server-side is therefore necessary to enable accurate decisions while preserving \textit{privacy} and \textit{regulatory compliance}.

%Across these domains, backend data enrichment is not an implementation detail but a structural necessity. Removing this step would force systems to rely on untrusted, stale, or incomplete user-provided inputs, or to embed rapidly changing state into models, both of which are impractical at scale while reducing the overall system's efficiency and increasing the response time. 
%Consequently, modern ML inference pipelines routinely incorporate server-side feature retrieval prior to neural network (NN) model execution, creating a setting 

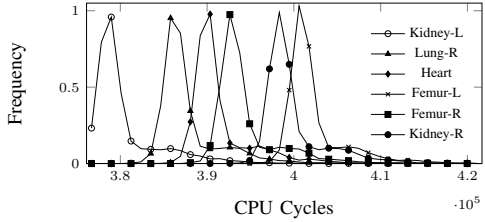
\begin{figure}[t]
\centering
\scalebox{0.75}{
\begin{tikzpicture}
\begin{axis}[
    xlabel={CPU Cycles},
    ylabel={Frequency},
    width=0.97\linewidth,
    height=0.5*\linewidth,
    xmin=376108, xmax=421678,
    ymin=0, ymax=1.05,
    ymajorgrids=false,
    legend style={
        font=\scriptsize,
        at={(0.98,0.9)},
        anchor=north east,
        legend columns=1,
        draw=none
    },
    xticklabel style={font=\scriptsize},
    yticklabel style={font=\scriptsize},
]

\pgfplotsset{
  tcurve/.style={
    mark repeat=2,
    mark size=2.4pt,
  }
}

% Kidney-L
\addplot+[tcurve, black,mark=o, color=black, mark size=1.5pt]
table[col sep=comma, x=bin_center, y expr=\thisrow{count}/3000]
{hist_kidney_left_bins40.csv};
\addlegendentry{Kidney-L}

% Lung-R
\addplot+[tcurve, black,mark=triangle*, mark options={draw=black, fill=black}, mark size=1.5pt]
table[col sep=comma, x=bin_center, y expr=\thisrow{count}/3000]
{hist_lung_right_bins40.csv};
\addlegendentry{Lung-R}

% Lung-L
% \addplot+[tcurve, brown,mark=square*]
% table[col sep=comma, x=bin_center, y expr=\thisrow{count}/3000]
% {hist_lung_left_bins40.csv};
% \addlegendentry{Lung-L}

% Heart
\addplot+[tcurve, black,mark=diamond*, mark options={draw=black, fill=black}, mark size=1.5pt]
table[col sep=comma, x=bin_center, y expr=\thisrow{count}/3000]
{hist_heart_bins40.csv};
\addlegendentry{Heart}

% Femur-L
\addplot+[tcurve, black, mark=x, mark options={draw=black, fill=black}, mark size=1.5pt]
table[col sep=comma, x=bin_center, y expr=\thisrow{count}/3000]
{hist_femur_left_bins40.csv};
\addlegendentry{Femur-L}

% Femur-R
\addplot+[tcurve, black,mark=square*,  mark options={draw=black, fill=black}, color=black, mark size=1.5pt]
table[col sep=comma, x=bin_center, y expr=\thisrow{count}/3000]
{hist_femur_right_bins40.csv};
\addlegendentry{Femur-R}

% % Kidney-R
% \addplot+[tcurve, dashed, black,mark=o, mark options={draw=green!50!black, fill=green!50!black}]
% table[col sep=comma, x=bin_center, y expr=\thisrow{count}/3000]
% {hist_kidney_right_bins40.csv};
% \addlegendentry{Kidney-R}
% Kidney-R (solid circle)
\addplot+[tcurve, solid, black, mark=*,
          mark options={draw=black, fill=black}, mark size=1.5pt]
table[col sep=comma, x=bin_center, y expr=\thisrow{count}/3000]
{hist_kidney_right_bins40.csv};
\addlegendentry{Kidney-R}

% Bladder
% \addplot+[tcurve, solid,yellow!60!black,mark=triangle*]
% table[col sep=comma, x=bin_center, y expr=\thisrow{count}/3000]
% {hist_bladder_bins40.csv};
% \addlegendentry{Bladder}

\end{axis}
\end{tikzpicture}
}
\caption{Normalized CPU-cycle inference-time %distributions 
for the MEDMNIST dataset, showing private diagnosis-dependent separations of up to 30K cycles (e.g., Kidney-L vs. Kidney-R) and revealing a strong sparsity-induced 
timing channel.}
\vspace{-1em}
%(pre-binned histograms; common binning) %\textcolor{red}{make the lines the same color as the markers, bring the legend box inside overlapping the plot area, on the top right, reduce the size of the markers, if you put markers you don't need colors, it can be black, this yellow especially is very bad, Ah my eyes hurt when I look at it.}.

\label{fig:cpu_cycles_norm}
\end{figure}

This paper explores the privacy of such systems where \textit{sensitive attributes influence inference behavior while remaining hidden from the API caller}, from the microarchitectural perspective for the first time, specifically by exploiting sparsity-driven zero-skipping optimizations in modern AI accelerators, which are widely implemented in today’s hardware. % and software stack. %co-design.
%\textcolor{blue}{
We show for the first time that inference time is distinct for different attribute values of user data.  Figure~\ref{fig:cpu_cycles_norm}
%{fig:Timing_based_histogram} 
shows the difference in inference time for a sensitive medical ML workload running on Intel's latest on-core AI  accelerator, AMX.
%}
%\textcolor{red}{is this AMX or GPU?}. 
%
We exploit this end-to-end correlation in \scheme{}, to leak user data values used in the training of the  %based on how fast hardware executes the 
ML applications. 
%, which were previously assumed to be secure. 

An attacker who knows a public or quasi-identifier such as a patient ID or SSN can submit a legitimate inference query associated with that identifier and observe its API response latency. Although sensitive attributes (e.g., whether the individual has a lung or heart condition) are retrieved internally via backend data enrichment and are never exposed through the API, they influence activation sparsity and hardware execution time. As a result, these hidden attributes can be inferred directly from the latency of the attacker’s own queries, without co-location, shared resources, or timing another user’s execution.

To mount \scheme{}, the adversary first profiles inference latency of it's own queries %conditioned on the hidden attribute e.g., Kidney-L vs. Kidney-R (Figure~\ref{fig:cpu_cycles_norm}) 
%(e.g., Major Depressive Disorder (MDD) vs.\ no-MDD) using representative inputs, 
and trains a secondary classifier on the collected timing traces together with the model’s returned label and any known (non-sensitive) attributes enabling inferences in systems where state-of-the-art attack strategies~\cite{Mehnaz2022Attribute} fundamentally fail. 
%At attack time, the adversary measures the end-to-end response latency of a query they issue and inputs the observed latency, predicted label,  and known attributes into the classifier to infer the private attribute, thereby recovering sensitive information via a microarchitectural timing channel.

These techniques such as attribute inference attacks  LOMIA and its confidence-based variant CSMIA~\cite{Mehnaz2022Attribute} is not effective here because (I) first they assume that the attacker can systematically \textit{flip} the sensitive attribute while observing ground-truth labels or confidence scores 
and, (II) they assume knowledge of the ground-truth output label.
These assumptions do not hold in modern production systems%-such as recommendation, fraud and risk scoring, credit underwriting, insurance decision support, and clinical decision support
 where sensitive attributes are retrieved internally via backend data enrichment and never exposed through the API. Consequently, the attacker cannot control the sensitive attribute or construct counterfactual queries that differ only in that attribute. 
 In contrast, our attack  does not require \textit{neither (I) flipping hidden attributes nor (II) access to ground-truth labels or (III) confidences}, thus, bypassing common defenses ~\cite{Mehnaz2022Attribute,%Jayaraman2022Attribute,
Gia2018Attriguard}. 

\medskip
\noindent\textbf{Contributions.} This paper makes the following contributions: 
\begin{itemize}
    % \item We present \scheme{}, the first hardware-level \emph{retrieval attribute-stealing attack} that leaks private %medical 
    % attributes solely through inference timing, bypassing confidence-masking and label-only defenses. 
    % % Exploiting Intel AMX’s sparsity-aware zero-skipping, 
    % % \scheme{} achieves %63.35\% accuracy in a two-class experiment (Procedures~26~vs.~93) and 
    % %  68.65\% average accuracy across ten classes on the Texas~100~Hospital dataset; an adversarial advantage of {+58.6~pp} over random guessing. \textcolor{red}{add new results.} 
    % % The two classes differ by {41,417~cycles}. 
    % % %(approximately 
    % % $10^{2}$--$10^{3}\!\times$ typical cache-attack timing gaps.
    % \scheme{} achieves high accuracy in identifying private attributes retrieved from the backend database which attacker does not have access on datasets like Texas100x, Census-19 and OrganAMNIST giving a huge adversarial advantage compared to the random guessing. We can leak the various private attributes among 10 Surgical procedure code operated on the patients with 60.98\%, 61.06\% and 70.3\% accuracy on backend hardwares like AVX, AMX and Nvidia A100 GPU respectively.
    \item We present \scheme{}, the first hardware-level \emph{backend enrichment attribute-stealing attack} that leaks private attributes exclusively through inference-time side effects, bypassing confidence-masking and label-only defenses.
    Across Texas100x, Census-19, and OrganAMNIST, \scheme{} enables an adversary to infer sensitive backend-retrieved attributes with high accuracy.
    For a 10-class surgical procedure inference task, the attack achieves 60.98\%, 61.06\%, and 70.3\% accuracy on AVX, AMX, and NVIDIA A100 backends, respectively, representing an adversarial advantage (AA) up to 98 pp over random guessing.

    \item We identify the \emph{root cause} of the leakage: data-dependent activation sparsity amplified by accelerator zero-skipping. Reproducing the attack on an NVIDIA~A100~GPU reveals operand-dependent latency %in \texttt{mma\_sync} 
    identical to AMX’s behavior and CPU AVX, confirming that the vulnerability generalizes across architectures, even with DVFS locked.

    \item We analyze mitigation strategies and show that %full 
    eliminating zero skipping as a defense %constant-time masking 
    imposes an%prohibitive 
    overhead of 25\%  and would double inference latency. We propose a padding defense that equalizes timing leakage across private attributes, with an average performance overhead of 7.24\% and no energy overhead.
    %We discuss a lightweight alternative \textcolor{red}{Farshad defense high level idea.}
    %- selective noise injection triggered only under suspicious or repetitive queries to reduce the privacy risk while preserving energy and performance advantage. %---which masks timing variations with under 3\% energy overhead, offering a practical path toward privacy-preserving accelerator design.
\end{itemize}

\noindent \textbf{Responsible Disclosure}. We disclosed this vulnerability to Intel and NVIDIA, who confirmed our findings while stating that they cannot guarantee privacy against such attacks due to the inherent timing advantage of the optimizations implemented in their products and recommended software-level defenses.

\vspace{-1em}
\section{Threat Model}

%\noindent\textbf{System Model.} 
%\textcolor{blue}{
We consider a label-only inference API that returns discrete predictions to the caller. During inference, the model internally retrieves exactly one sensitive attribute from a backend source (e.g., a feature store or database) and incorporates it into the inference pipeline. This sensitive attribute is never exposed through the API and is not part of the client-provided input.
%}

% feature-set retrieval and enrichment system which retrieves private, backend-stored features at inference time (from Cassandra / feature store),
% joins them with request context,
% constructs a hidden feature vector,
% and then runs a model whose execution depends on those retrieved features.
\noindent\textbf{Attacker.} 
The attacker in our setting is the entity issuing inference queries to the system, not a separate victim whose execution is being timed. The attacker interacts with the service as a legitimate remote API user, similar to prior black-box settings~\cite{Mehnaz2022Attribute}, issues standard inference requests, and measures the end-to-end response latency of \emph{their own} queries. No co-location, shared hardware, or timing of other users’ executions is required. The attacker has black-box access only, with no access to model weights, gradients, logits, or confidence scores.
\scheme{} exploits class-dependent inference-time gaps of 20–30K CPU cycles (up to 40K), which are roughly two orders of magnitude larger than the 50–400 cycle differences used in remote attacks such as HertzBleed, NetSpectre, and GateBleed. As a result, the attack is robust in fully remote settings without co-location or shared resources.

\noindent\textbf{Target.} 
The attacker’s goal is to infer a sensitive attribute of training data that is never directly present in the client’s input but is fetched internally from a backend or feature store during inference (e.g., in retrieval-augmented generation pipelines, offline-to-online feature pipelines, or hospital EHR retrieval). Given non-sensitive inputs, the returned label, and the observed inference latency, the attacker aims to infer the hidden backend attribute of a target individual among \(k\) sensitive classes.
%\noindent\textbf{Remote Timing Assumptions.} 
All measurements are performed remotely by timing the API responses to attacker-issued queries. The timing signal arises solely from the internal execution of the backend-enriched inference pipeline on hardware accelerators and propagates to the observable end-to-end latency. Because the attacker controls the queries they submit, repeated queries used for profiling do not require access to or interaction with any other user.%\noindent\textbf{Backend Data Enrichment and Realism.} 
% Backend data enrichment is a standard and necessary design in modern healthcare and insurance decision support systems. 
%\noindent\textbf{ Example.} 
An example is intelligence/military settings, in an online decision service where the request contains only an identifier and minimal context, but the model’s prediction requires backend-enriched features pulled from protected databases at inference time such as a target-prioritization API where a client submits a target ID plus current location/time, and the system retrieves classified context such as prior sightings, watchlist status, threat tier, and recent activity from intelligence stores, then runs a ranking model to decide which targets to task next. 

 \vspace{-1em}
\section{\scheme{}}
% To mount \scheme{}, the adversary first profiles inference latency conditioned on the hidden attribute (e.g., Major Depressive Disorder (MDD) vs.\ no-MDD) using representative inputs, and trains a secondary classifier on the collected timing traces together with the model’s returned label and any known (non-sensitive) attributes. At attack time, the adversary measures the end-to-end response latency of a query they issue and inputs the observed latency, predicted label, and known attributes into the classifier to infer the private attribute. 
%, thereby recovering sensitive information via a microarchitectural timing channel.

\paragraph*{Step 1: Profiling retrieval-induced timing.}
Using auxiliary or synthetic identifiers, the attacker repeatedly queries the system and records inference latency. Each identifier triggers the retrieval of a hidden set of backend attributes, which induces a characteristic activation sparsity pattern and, consequently, a distinct execution-time profile. Over repeated measurements, these profiles form stable timing fingerprints associated with different backend attribute values.

\paragraph*{Step 2: Learning timing clusters for backend attributes.}
The attacker clusters the collected timing profiles to identify groups corresponding to different hidden attribute configurations (e.g., high-risk vs.\ low-risk, genetic marker present vs.\ absent). When a small labeled subset of identifiers is available through partial knowledge or leakage, the attacker can assign semantic labels to these clusters. To scale to multi-class settings and improve robustness, the attacker trains a Gradient Boosted Decision Tree (GBDT) classifier using inference time, non-sensitive inputs, and predicted labels as features. The attacker is assumed to have access to an auxiliary dataset drawn from the same population but disjoint from the victim model’s training set.

\paragraph*{Step 3: Retrieved data inference.}
Once the timing clusters are established, the attacker observes the latency of a target query (e.g., associated with a specific identifier) and assigns it to the closest cluster, thereby inferring properties of the retrieved backend data. Because backend retrieval alters activation sparsity and sparsity directly affects execution time on zero-skipping accelerators, this attack bypasses output/confidence masking-based defenses. 

\vspace{-1em}
\section{Security Analysis}
%\subsection{Model \& dataset}

\noindent\textbf{Models and Leakage Across Architectures.} 
We evaluate \scheme{} across \textbf{DNNs (tabular inference)},\textbf{ CNNs (image-based inference)}, and \textbf{hybrid CNN--MLP pipelines}. On NVIDIA A100, DNNs exhibit the strongest leakage, achieving up to 70.3\% accuracy, F1 up to 98.7, and adversarial advantage (AA) up to +98.9 pp (percentage points) on Texas-100X. In contrast, CNN-based pipelines on OrganAMNIST leak less on average, with 25.09\% accuracy, F1 between 14.7--43.3, and AA up to +51.7 pp, indicating that dense visual representations attenuate but do not eliminate timing leakage. Leakage persists across all model families tested, demonstrating architectural generality.

 \begin{table}[!htbp]
\centering
\small
\renewcommand{\arraystretch}{0.9}
\setlength{\tabcolsep}{3pt}  % tighter columns
\scalebox{0.75}{
\begin{tabular}{l l c c c c}
\toprule
\textbf{Dataset} & \textbf{Attribute} & \textbf{Class} &
\textbf{Accuracy} & \textbf{F1} & \textbf{AA} \\
\midrule
OrganAMNIST & CT (Heart)         & 11 & 56.08 & 17.62 & +51.69 \\
OrganAMNIST & CT (Femur-L)       & 11 & 50.37 & 21.48 & +45.41 \\
OrganAMNIST & CT (Femur-R)       & 11 & 48.71 & 20.37 & +43.58 \\
OrganAMNIST & CT (Kidney-L)      & 11 & 45.97 & 43.26 & +40.57 \\
OrganAMNIST & CT (Kidney-R)      & 11 & 43.27 & 41.34 & +37.60 \\
OrganAMNIST & CT (Liver)         & 11 & 43.12 & 43.06 & +37.43 \\
OrganAMNIST & CT (Bladder)       & 11 & 11.60 & 14.71 &  +2.76 \\
% OrganAMNIST & CT (Lung-R)        & 11 &  9.09 &  3.23 &  0.00  \\
% OrganAMNIST & CT (Lung-L)        & 11 &  9.09 &  1.49 &  0.00  \\
% OrganAMNIST & CT (Pancreas)      & 11 &  9.09 &  0.87 &  0.00  \\
% OrganAMNIST & CT (Spleen)        & 11 &  9.09 &  0.85 &  0.00  \\
Texas-100X  & Surg: C-sec        & 10 & 98.98 & 98.71 & +98.87 \\
Texas-100X  & Surg: Prostate     & 10 & 97.77 & 77.23 & +97.52 \\
Texas-100X  & Surg: Appendix     & 10 & 81.75 & 88.42 & +79.72 \\
Texas-100X  & Surg: Brain        & 10 & 74.27 & 82.21 & +71.41 \\
Texas-100X  & Surg: Gallbladder  & 10 & 74.00 & 65.58 & +71.11 \\
Texas-100X  & Surg: Abdomen      & 10 & 52.23 & 37.56 & +46.92 \\
Texas-100X  & Surg: Derm.        & 10 & 30.48 & 32.16 & +22.76 \\
Texas-100X  & Surg: Blood Ves.   & 10 & 23.83 & 25.19 & +15.37 \\
% Texas-100X  & Surg: Chest        & 10 &  0.00 &  0.00 &  0.00  \\
% Texas-100X  & Surg: Kidney       & 10 &  0.00 &  0.00 &  0.00  \\
Census-19 & Marital: Married        & 5 & 37.66 & 0.5 & +17.66 \\
Census-19 & Marital: Widowed        & 5 & 60.32 & 0.12 & +40.32 \\
Census-19 & Marital: Divorced       & 5 & 23.71 & 0.19 & +3.71 \\
Census-19 & Marital: Separated      & 5 & 33.12 & 0.04 & +13.12 \\
Census-19 & Marital: Never Married  & 5 & 65.59 & 0.70 & +45.59 \\

\bottomrule
\end{tabular}
}
\caption{Inference accuracy (Acc.), weighted F1 (W-F1), 
and adversarial advantage (AA) across all scenarios} % on Nvidia A100 GPU. }
\vspace{-1em}
\label{tab:leakage_all_summary}
\end{table}
\noindent\textbf{Datasets and Attribute Sensitivity.} 
Across datasets, \textbf{Texas-100X} exhibits the highest average leakage, followed by \textbf{Census-19} and \textbf{OrganAMNIST}. On \textbf{Texas-100X}, FeatureBleed achieves an average accuracy of 70.3\% over 10 classes with AA exceeding +70 pp for multiple surgical procedures. On \textbf{Census-19}, the attack achieves 45.24\% average accuracy over 5 marital-status classes with AA up to +45.6 pp. On \textbf{OrganAMNIST}, average accuracy is lower (25.09\% over 11 classes), yet remains well above random guessing, confirming leakage across data modalities.

\noindent\textbf{Model Size Impact.}
% \textcolor{red}{this should be AA or accuracy or success rate something like this. Fix this analysis and end with a conclusion, you had before about width vs depth etc..} 
% We performed a systematic model-size-scaling of 336 to 1053696 parameters for DNN (Table ~\ref{tab:channel_strength_params}), showing increasing model size only strengthens our attack with Cohen’s D spanning from 0.0115 to 4.7800 and our attack accuracy using a simple threshold classifier to classify across 2 classes is proportional to the Cohen's D. The detailed impact of varying width (8 to 512) versus depth (2 to 7) is also seen in the table. We see that Cohen's D and the attack accuracy increases linearly as we increase the width keeping the depth constant to 4. But in case of varying the depth, we that Cohen's D and Attack accuracy increases until depth of 5 and decreases as depth rises indicating that sparsity saturation has been reach.
We perform a systematic model-size scaling study, ranging from 336 to 1,053,696 parameters, for fully connected DNNs (Table~\ref{tab:channel_strength_params}). Our results show that increasing model size consistently strengthens the attack. Cohen’s 
$d$ measures the standardized separation between two distributions, capturing how distinguishable they are relative to noise.%, with Cohen’s $d$ increasing from 0.0115 to 4.7800. Correspondingly, the attack accuracy of a simple threshold classifier for binary classification closely tracks Cohen’s $d$, confirming that effect size is a reliable predictor of attack success.
%The table further disentangles the impact of width (8–512) and depth (2–7). 
When the width is increased while keeping the depth fixed, %at 4, 
both Cohen’s $d$ and attack accuracy grow linearly, indicating that wider models amplify the exploitable leakage signal. 
In contrast, when depth is varied for a fixed width, Cohen’s $d$ and attack accuracy increase up to a depth of five. 
%5 but then decline for deeper models, suggesting that sparsity saturation is reached beyond this point, limiting further leakage amplification.

%We experimented with different model introducing variations in model width and depth as shown in Table ~\ref{tab:channel_strength_params}. Cohen’s D (effect size) is used to quantify how far apart two timing distributions are. 
%Cohen's D between $0.2$ and $0.5$ is quantified as small leakage, between $0.5$ and $0.8$ is classified as medium leakage and bigger than $0.8$ is classified as huge leakge. 
%Here we see that Cohen's d is consistently yielding large effect size accross various MLP architectures thus indicating that phenomenon consistently persists on different model families rather than arising from a single "pathological network". 

\begin{table}[!htbp]
\centering
\small
\scalebox{0.75}{
\begin{tabular}{r r r r r r}
\toprule
\textbf{Width} & \textbf{Depth} & \textbf{\# Params} & \textbf{\# Activations} & \textbf{Cohen’s $d$} & \textbf{Accuracy} \\
\midrule
8   & 4 & 336        & 288        & 0.0115 & 50.14\% \\
16  & 4 & 1,184      & 1,088      & 0.0271 & 50.85\% \\
32  & 4 & 4,416      & 4,224      & 0.1045 & 51.35\% \\
64  & 4 & 17,024     & 16,640     & 0.1998 & 53.05\% \\
128 & 2 & 34,048     & 33,024     & 0.0500 & 50.93\% \\
128 & 3 & 50,432     & 49,536     & 0.2800 & 53.96\% \\
128 & 4 & 66,816     & 66,048     & 0.5552 & 63.42\% \\
128 & 5 & 83,200     & 82,560     & 0.9500 & 70.37\% \\
128 & 6 & 99,584     & 99,072     & 0.7800 & 69.17\% \\
128 & 7 & 116,352    & 115,584    & 0.65   & 67.87\% \\
256 & 4 & 264,704    & 263,198    & 0.4767 & 61.61\% \\
512 & 4 & 1,053,696  & 1,050,624  & 4.7800 & 99.37\% \\
\bottomrule
\end{tabular}
}
\caption{Channel strength for various  model sizes.
}
%Parameter count denotes the total number of weights assuming input dimension equals width and a fixed output dimension of 10.Activation count denotes the total number of neuron activations executed per inference.}
\label{tab:channel_strength_params}
\end{table}
% \noindent\textbf{Comparative Effectiveness.} 
% Overall, adversarial advantage correlates strongly with activation sparsity and model depth. DNNs operating on backend-enriched tabular data leak the most (AA up to \textbf{+98.9 pp}), while CNN-based pipelines leak less but still exhibit substantial advantage over random guessing (AA up to \textbf{+51.7 pp}). Weighted F1 scores follow the same trend, indicating that leakage is not driven by class imbalance but by systematic timing separability induced by sparsity-driven execution.

% \noindent\textbf{Texas100x} We train a three-layer perceptron (MLP) with layer widths \{256,128,64, ReLU\}

\noindent\textbf{Hardware Accelerator Generality.}
We evaluate FeatureBleed across three widely deployed hardware backends such as\textbf{Intel AVX CPUs}, \textbf{Intel AMX accelerators}, and \textbf{NVIDIA A100 GPUs} as summarized in Table~\ref{tab:diff_hardware_leak}. The attack consistently infers multiple sensitive attributes across all platforms, demonstrating that the leakage is not tied to a specific microarchitecture. On Intel AVX and AMX, FeatureBleed achieves comparable accuracy for surgical procedure codes (60.98\% and 61.06\%, respectively), while the A100 exhibits higher leakage, reaching 70.30\% accuracy. Similar trends hold across other attributes (sex, ethnicity, race, and source of admission), with GPUs consistently yielding the highest inference accuracy. These results indicate that sparsity-driven timing leakage generalizes across CPU and GPU accelerators, and that more aggressive sparsity optimizations amplify the attack.

% This experiments is done on three different hardware backends like Intel AVX, Intel AMX and Nvidia A100 GPU as shown in Table~\ref{tab:diff_hardware_leak}.

\begin{table}[t]
\centering
\small
\setlength{\tabcolsep}{3.5pt}
\renewcommand{\arraystretch}{1.1}
\label{tab:platform_attribute_leakage}
\scalebox{0.85}{
\begin{tabular}{l c c c c c}
\toprule
\textbf{Platform} & \textbf{Proc.\ Code} & \textbf{Sex} & \textbf{Ethnicity} & \textbf{Race} & \textbf{\shortstack{Admission}} \\
\midrule
AVX & 60.98\% & 58.19\% & 74.57\% & 23.15\% & 41.40\% \\
AMX  & 61.06\% & 61.63\% & 74.86\% & 20.89\% & 43.77\% \\
GPU  & 70.30\% & 66.25\% & 80.07\% & 43.57\% & 50.87\% \\
\bottomrule
\end{tabular}
}
\caption{ Inference accuracy across different accelerators.}
\label{tab:diff_hardware_leak}
\end{table}

\vspace{-1em}
\section{Root Cause Analysis and Mitigation}

 We experimentally confirm that Intel AMX, the on-core accelerator in recent Xeon processors, exhibits zero-skipping behavior where its inference latency scales with operand sparsity, which in turn correlates with specific data attributes. Intel AMX has a special matrix multiplication instruction \texttt{TDPBSSD} 
%(Tile Dot Product of Signed Bytes with Dword) 
which we use for ML inference in this paper.
This input-dependent latency forms the hardware timing channel that \scheme{} exploits to leak sensitive attributes.
We reproduced \scheme{} leakage on the NVIDIA A100 GPU to confirm that zero-skipping optimizations extend beyond Intel AMX. \texttt{mma\_sync} is a Tensor Core matrix multiply–accumulate instruction on NVIDIA GPUs. It performs warp-level matrix multiplication on Tensor Cores and is used internally by CUDA kernels for ML inference (e.g., GEMM), which exhibits a mean operand-dependent timing difference of 10,000 cycles.
This generalizes leakage beyond Intel AMX, though it is weaker due to off-chip data movement. 
%
% On-core AMX leakage is observable within approximately \(10{,}000\) iterations, 
% whereas the GPU requires about \(1{,}000{,}000\) iterations to achieve comparable distinguishability.

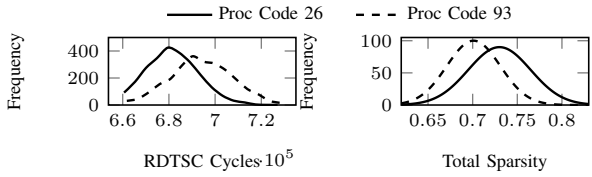
\begin{figure}[t]
\centering
\begin{tikzpicture}

% ---------- shared legend (top) ----------
\begin{axis}[
    hide axis,
    width=\linewidth,
    height=1.2cm,          % <-- NOT 0pt
    scale only axis,
    xmin=0, xmax=1, ymin=0, ymax=1, % dummy limits
    legend columns=2,
    legend cell align=left,
    legend style={
        at={(0.35,1)},     % <-- place INSIDE this tiny axis
        anchor=center,
        draw=none,
        fill=white,
        fill opacity=0.85,
        text opacity=1,
        font=\scriptsize,
        /tikz/every even column/.append style={column sep=1.0em}
    },
]
\addlegendimage{black, solid, line width=0.9pt}
\addlegendentry{Proc Code 26}
\addlegendimage{black, dashed, line width=0.9pt}
\addlegendentry{Proc Code 93}
\end{axis}

% ---------- the two plots ----------
\begin{groupplot}[
    group style={
        group size=2 by 1,
        horizontal sep=1.4cm,
    },
    width=0.46\linewidth,
    height=0.28\linewidth, % short but readable
    tick label style={font=\scriptsize},
    label style={font=\scriptsize},
    axis line style={line width=0.6pt},
    tick style={line width=0.5pt},
    grid=none,
]

% (a) Timing histogram (smooth + not packed)
\nextgroupplot[
    xlabel={RDTSC Cycles},
    ylabel={Frequency},
    ymin=0,
    ymax=500,
    % if you want, you can also set xmin/xmax to zoom in
    % xmin=660000, xmax=730000,
]

% Proc 26: smooth line (no markers) + fewer points drawn (packed fix)
\addplot+[
    black, solid, line width=0.9pt,
    no marks,
    smooth,
    each nth point=4
]
table[
    col sep=comma,
    x=x, y=freq,
    restrict expr to domain={\thisrow{type}}{26:26}
]{Diagrams/proc26_vs_93_hist.csv};

% Proc 93
\addplot+[
    black, dashed, line width=0.9pt,
    no marks,
    smooth,
    each nth point=4
]
table[
    col sep=comma,
    x=x, y=freq,
    restrict expr to domain={\thisrow{type}}{93:93}
]{Diagrams/proc26_vs_93_hist.csv};

% (b) Sparsity distribution
\nextgroupplot[
    xlabel={Total Sparsity},
    ylabel={Frequency},
    xmin=0.62, xmax=0.83,
    ymin=0, ymax=105,
]

\addplot[smooth, line width=0.9pt, black, domain=0.62:0.83, samples=200]
{90*exp(-0.5*((x-0.730)/0.035)^2)};

\addplot[smooth, line width=0.9pt, black, dashed, domain=0.62:0.83, samples=200]
{100*exp(-0.5*((x-0.701)/0.030)^2)};

\end{groupplot}

\end{tikzpicture}

\caption{Timing and sparsity distributions for procedure codes 26 and 93.  Left: inference timing histogram. Right: total sparsity distribution.}
\vspace{-1em}
\label{fig:timing_sparsity_26_93}
\end{figure}

We measure attribute-dependent activation sparsity %on Intel AMX 
and observe that each procedure has a different end-to-end mean sparsity. For example, the procedure code 26 has %induces a mean sparsity of \textbf{0.730}, compared to \textbf{0.701} for procedure code 93, corresponding to a \textbf{$\approx$
4\% higher sparsity than the procedure code 93; therefore samples with Procedure code 26 take less time to execute than sample of Procedure code 93 as shown in Figure \ref{fig:timing_sparsity_26_93}. Although the distributions partially overlap for some attributes, the sparsity difference is systematic and stable across runs and different datasets and the timing distribution  of some attributes completely separates (see Figure \ref{fig:cpu_cycles_norm}) resulting in up to 98\% accuracy (see table~\ref{tab:leakage_all_summary}) for example Surgical Procedure Code of C-Section in Texas100x, or heart vs kidney in MEDMNIST dataset.
%Because Intel AMX employs sparsity-driven zero-skipping, this attribute-dependent sparsity gap directly translates into measurable inference-time differences, confirming that backend-retrieved attributes induce distinct and exploitable execution behavior.

Thus, sensitive backend retrieved data with varied sparsity creates a novel, observable side-channel. We show this correlation between class vs. sparsity and sparsity vs. time%~\cite{Dizani2025Thor}%\textcolor{blue}{TODO this figure.}
, leads to the exploited correlation between
class vs. time %(Figure~\ref{fig:Timing_based_histogram}) \textcolor{red}{I guess we can remove this figure too and refer to figure 1 when needed}, 
and also confirms the presence of zero skipping with energy analysis.
Thus, we conclude that the root cause of this leakage is two-fold: (1) data features correlate with sparsity due to representation learning in NNs, and ReLU further amplifies this sparsity %(Figure~\ref{fig:cleaned-cohen-activation}) 
by zeroing negative values, and (2) sparsity correlates with execution time due to zero skipping in accelerators in hardware. 
%Thus, the varying number of zeros during neural network computations impacts execution time on hardware accelerators such as Intel’s Advanced Matrix Extensions (AMX). We show that the inference optimized for Intel AMX is faster when a higher proportion of operands are zero. 
Therefore, variations in runtime can inadvertently leak information about the retrieved backend data through side-channel timing analysis, revealing sensitive attributes of the data.

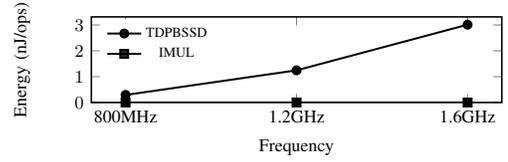
\begin{figure}[!t]
\centering
\scalebox{0.75}{
\begin{tikzpicture}
\begin{axis}[
    width=\linewidth,
    height=0.35\linewidth,
    xlabel={Frequency},
    ylabel={Energy (nJ/ops)},
    font=\small,
    thick,
    ymin=0,
    grid=none,
    xtick=data,
    symbolic x coords={800MHz,1.2GHz,1.6GHz},
    legend style={
        at={(0.02,0.98)},
        anchor=north west,
        draw=none,
        fill=white,
        fill opacity=0.85,
        text opacity=1,
        font=\scriptsize
    },
]

% ---- TDPBSSD ----
\addplot+[
    black,
    mark=*,
     mark options={draw=black, fill=black},
    line width=1pt
]
coordinates {
    (800MHz,0.292057)
    (1.2GHz,1.245076)
    (1.6GHz,3.011822)
};

% ---- IMUL ----
\addplot+[
    black,
    mark=square*,
     mark options={draw=black, fill=black},
    line width=1pt
]
coordinates {
    (800MHz,0)
    (1.2GHz,0)
    (1.6GHz,0)
};

\legend{TDPBSSD, IMUL}

\end{axis}
\end{tikzpicture}
}
\caption{Energy gap between operands 0 and 255 across fixed CPU frequencies. AMX TDPBSSD remains operand-dependent even when DVFS is locked.}
\vspace{-1em}
\label{fig:Mean_difference_frequency}
\end{figure}

Our results show disabling TurboBoost/DVFS prevents frequency-scaling leaks~\cite{Wang2022Hertzbleed} but not \scheme{}; the energy gap between 0 and 255 operands (Fig.~\ref{fig:Mean_difference_frequency}) grows even under frequency lock, confirming a zero-skipping effect. Therefore, if we disable zero-skipping,
%even if no such software control exists, 
the average energy of TDPBSSD operations with all-zero operands is 12 nJ versus all-one operands is 15 nJ, yielding an increase of $[(15nJ - 12nJ)/12nJ]*100 \approx 25\%$ energy per operation. Defenses that clip or randomize confidences~\cite{Mehnaz2022Attribute, Gia2018Attriguard} %,ShokriSS16} 
remove statistical cues, but not timing leakage; \scheme{} remains effective since latency still correlates with sparsity.
%Software-level constant-time coding cannot eliminate the leakage 
% Because the timing bias originates inside the accelerator, only rebuilding the hardware by eliminating the zero-skipping optimization, or not using ReLu in SW would close the channel, but such full masking doubles inference latency and raises energy by $\approx$25\%. 
%We estimate this number by subtracting the average energy taken by TDBPSSD operation if all the operands are zeros(12nJ) from the TDBPSSD operation if all the operands were ones(15nJ) $[(15nJ - 12nJ)/12nJ]*100 \approx 25\%$. We find no software control to disable zero-skipping. 
%

% \begin{figure}[!t]
%     \centering
%     % \vspace{-2em} 
%     \includegraphics[width=0.6\linewidth]{images/frequency_comparison.pdf}
%     \caption{Energy gap between operands 0 and 255 across fixed CPU frequencies.  
%     AMX \texttt{TDPBSSD} remains operand-dependent even when DVFS is locked. \textcolor{red}{figures don't need title, remove the title, we have caption, make the plot wider and shorter to save space, }}
%     \vspace{-1em}
%     \label{fig:Mean_difference_frequency}
% \end{figure}

%Disabling TurboBoost or DVFS prevents frequency-scaling leaks (e.g., Hertzbleed~\cite{Wang2022Hertzbleed}) but not \scheme{}.  In our measurements (Fig.~\ref{fig:Mean_difference_frequency}), the energy gap between all-zero and all-255 operands \emph{increases} with frequency even under lock, proving DVFS is not the cause and confirming a microarchitectural zero-skipping effect.
%

% 
 % Thus, we propose a  padding defense in which the API withholds the response until a fixed budget equal to the worst-case end-to-end inference time elapses. This masks timing variations without performing extra computation; the additional energy is largely limited to idle power during the padding interval.

%\noindent{\textbf{Software-Level Defense.}}
We propose a padding defense in which the API withholds the response until a fixed budget equal to the worst-case end-to-end inference time elapses, masking timing variations without performing extra computation. The additional energy is largely limited to idle power during the padding interval. Average performance overhead across three hardware on Texas-100X (80–90\% sparsity) is 7.24\%.

\vspace{-0.5em}\section{Conclusion}
We show that microarchitectural %performance and power 
optimizations in AI accelerators, such as %sparsity-driven 
zero-skipping, create new timing channels that can leak private backend-retrieved attributes in %enriched 
ML pipelines. Future work should pursue hardware--software co-design defenses that mitigate such rising leakages while preserving the performance and energy benefits of accelerator optimizations.

% \vspace{-1em}
\vspace{-0.65em}
\section*{Acknowledgment}
% The authors thank anonymous reviewers for their helpful comments and feedback. 
This work was supported by Semiconductor Research Corporation (SRC) contract \#2025-HW-3306 and Intel Labs.

\vspace{-0.5em}

\begin{spacing}{0.92}
   \bibliographystyle{IEEEtranS}
\bibliography{refs}

@inproceedings{de2024hopsworks,
  title={The hopsworks feature store for machine learning},
  author={de la R{\'u}a Mart{\'\i}nez, Javier and Buso, Fabio and Kouzoupis, Antonios and Ormenisan, Alexandru A and Niazi, Salman and Bzhalava, Davit and Mak, Kenneth and Jouffrey, Victor and Ronstr{\"o}m, Mikael and Cunningham, Raymond and others},
  booktitle={Companion of the 2024 International Conference on Management of Data},
  pages={135--147},
  year={2024}
}

@inproceedings {Mehnaz2022Attribute,
author = {Shagufta Mehnaz and Sayanton V. Dibbo and Ehsanul Kabir and Ninghui Li and Elisa Bertino},
title = {Are Your Sensitive Attributes Private? Novel Model Inversion Attribute Inference Attacks on Classification Models},
booktitle = {31st USENIX Security Symposium (USENIX Security 22)},
year = {2022},
isbn = {978-1-939133-31-1},
address = {Boston, MA},
pages = {4579--4596},
url = {https://www.usenix.org/conference/usenixsecurity22/presentation/mehnaz},
publisher = {USENIX Association},
month = aug
}

@inproceedings{Wang2022Hertzbleed,
  title={Hertzbleed: Turning power side-channel attacks into remote timing attacks on x86},
  author={Wang, Yu and others},
  booktitle={USENIX Security},
  year={2022}
}

@inproceedings {Gia2018Attriguard,
author = {Jinyuan Jia and Neil Zhenqiang Gong},
title = {{AttriGuard}: A Practical Defense Against Attribute Inference Attacks via Adversarial Machine Learning},
booktitle = {27th USENIX Security Symposium (USENIX Security 18)},
year = {2018},
isbn = {978-1-939133-04-5},
address = {Baltimore, MD},
pages = {513--529},
url = {https://www.usenix.org/conference/usenixsecurity18/presentation/jia-jinyuan},
publisher = {USENIX Association},
month = aug
}

@misc{nvidia2023featurestore,
  author={NVIDIA},
  title={Offline-to-Online Feature Storage for Real-Time Recommendation Systems},
  howpublished={\url{https://developer.nvidia.com/blog/offline-to-online-feature-storage-for-real-time-recommendation-systems-with-nvidia-merlin/}},
  year={2023}
}

@article{ML_feature_stores, author = {Orr, Laurel and Sanyal, Atindriyo and Ling, Xiao and Goel, Karan and Leszczynski, Megan}, title = {Managing ML pipelines: feature stores and the coming wave of embedding ecosystems}, year = {2021}, issue_date = {July 2021}, publisher = {VLDB Endowment}, volume = {14}, number = {12}, issn = {2150-8097}, url = {https://doi-org.prox.lib.ncsu.edu/10.14778/3476311.3476402}, doi = {10.14778/3476311.3476402}, numpages = {4} }

@misc{feast_feature_store_2025,
  title        = {Feast: The Open Source Feature Store for Machine Learning},
  author       = {{Feast Project Contributors}},
  howpublished = {\url{https://feast.dev}},
  year         = {2025},
  note         = {Accessed: 2025-12-29},
  institution  = {LF AI \& Data / Feast Community},
  url          = {https://feast.dev}
}

@misc{uber2024predictive_to_generative,
  title        = {From Predictive to Generative – How Michelangelo Accelerates Uber’s AI Journey},
  author       = {{Uber Engineering}},
  year         = {2024},
  month        = may,
  day          = {2},
  howpublished = {\url{https://www.uber.com/blog/from-predictive-to-generative-ai/}},
  note         = {Accessed: 2025-12-30}
}
  
\end{spacing}

\vfill
\end{spacing}
\end{document}